\documentclass[pra,twocolumn,showkeys,showpacs]{revtex4}
\usepackage{epsf}

\def\llangle{\langle\!\langle}
\def\rrangle{\rangle\!\rangle}

\begin{document}

\title{Weak Measurements, Quantum State Collapse and the Born Rule}
\author{Apoorva Patel}
\email{adpatel@cts.iisc.ernet.in}
\author{Parveen Kumar}
\email{parveenkumar@cts.iisc.ernet.in}
\affiliation{Centre for High Energy Physics,
    Indian Institute of Science, Bangalore 560012, India}

\date{7 July 2017}

\begin{abstract}
Projective measurement is used as a fundamental axiom in quantum mechanics,
even though it is discontinuous and cannot predict which measured operator
eigenstate will be observed in which experimental run. The probabilistic
Born rule gives it an ensemble interpretation, predicting proportions of
various outcomes over many experimental runs. Understanding gradual weak
measurements requires replacing this scenario with a dynamical evolution
equation for the collapse of the quantum state in individual experimental
runs. We revisit the quantum trajectory framework that models quantum
measurement as a continuous nonlinear stochastic process. We describe
the ensemble of quantum trajectories as noise fluctuations on top of
geodesics that attract the quantum state towards the measured operator
eigenstates. In this effective theory framework for the ensemble of
quantum trajectories, the measurement interaction can be specific to
each system-apparatus pair---a context necessary for understanding weak
measurements. Also in this framework, the constraint to reproduce projective
measurement as per the Born rule in the appropriate limit, requires that
the magnitudes of the noise and the attraction are precisely related, in
a manner reminiscent of the fluctuation-dissipation relation. This relation
implies that both the noise and the attraction have a common origin in the
underlying measurement interaction between the system and the apparatus.
We analyse the quantum trajectory ensemble for the scenarios of quantum
diffusion and binary quantum jump, and show that the ensemble distribution
is completely determined in terms of a single evolution parameter. This
trajectory ensemble distribution can be tested in weak measurement
experiments. We also comment on how the required noise may arise in the
measuring apparatus.
\end{abstract}



\maketitle

\section{Background}

The axiomatic formulation of quantum mechanics has two distinct dynamical
mechanisms for evolving a state. One is unitary evolution, specified by
the Schr\"odinger equation:
\begin{equation}
\label{schrevol}
i{d \over dt}|\psi\rangle = H|\psi\rangle ~,~~
i{d \over dt}\rho = [H,\rho] ~.
\end{equation}
It is continuous, reversible and deterministic. The other is the von
Neumann projective measurement, which gives one of the eigenvalues of
the measured observable as the measurement outcome and collapses the
state to the corresponding eigenvector. With $P_i$ denoting the
projection operator for the eigenstate $|i\rangle$,
\begin{eqnarray}
|\psi\rangle \longrightarrow P_i |\psi\rangle / |P_i |\psi\rangle| , &&\\
P_i = P_i^\dagger ,~~ P_i P_j = P_i \delta_{ij} ,~~ && \sum_i P_i = I .
\end{eqnarray}
This change is discontinuous, irreversible and probabilistic in the
choice of ``$i$". It is consistent on repetition, i.e. a second
measurement of the same observable on the same system gives the same
result as the first one.

Both these evolutions, not withstanding their dissimilar properties,
take pure states to pure states. They have been experimentally verified
so well that they are accepted as basic axioms in the theoretical
formulation of quantum mechanics. Nonetheless, the formulation misses
something: {\it While the complete set of orthogonal projection operators
$\{P_i\}$ is fixed by the measured observable, only one ``$i$" occurs in
a particular experimental run, and there is no prediction for which ``$i$"
that would be.}
 
What appears instead in the formulation is the probabilistic Born rule,
requiring an ensemble interpretation for the outcomes. Measurement of an
observable on a collection of identically prepared quantum states gives:
\begin{equation}
\label{Bornrule}
prob(i) = \langle\psi|P_i|\psi\rangle = Tr(P_i\rho) ~,~~
\rho \longrightarrow \sum_i P_i \rho P_i ~.
\end{equation}
This rule evolves pure states to mixed states. All predicted quantities
are expectation values obtained as averages over many experimental runs.
The mixed state also necessitates a density matrix description, instead
of a ray in the Hilbert space description for a pure state.

\subsection{Environmental Decoherence}

Over the years, many attempts have been made to combine the two distinct
quantum evolution rules in a single framework. Although the problem of
which ``$i$" will occur in which experimental run has remained unsolved,
understanding of the ``ensemble evolution" of a quantum system has been
achieved in the framework of environmental decoherence. This framework
assumes that both the system and its environment (which includes the
measuring apparatus) are governed by the same set of basic quantum rules.
The essential difference between the system and the environment is that the
degrees of freedom of the system are observed while those of the environment
are not. Consequently, all the unobserved degrees of freedom need to be
``summed over" to determine how the remaining observed degrees of freedom
evolve.

Interactions between the system and its environment, with a unitary
evolution for the whole universe, entangles the observed system degrees
of freedom with the unobserved environmental degrees of freedom. The extent
of this entanglement can be controlled somewhat, by designing experiments
where the system mostly interacts with the measuring apparatus and has
little direct interaction with the rest of the environment. When the
unobserved degrees of freedom are summed over, a pure but entangled state
for the whole universe reduces to a mixed density matrix for the system:
\begin{equation}
|\psi\rangle_{SE} \longrightarrow U_{SE} |\psi\rangle_{SE} ~,~~
\rho_S = Tr_E(\rho_{SE}) ~,~~ \rho_S^2 \ne \rho_S ~.
\end{equation}
In general, the evolution of a reduced density matrix is linear,
Hermiticity preserving, trace preserving and positive, but not unitary.
Such a superoperator evolution can be expressed in the Kraus decomposition
form:
\begin{eqnarray}
\label{krausdecomp}
&& \rho_S \longrightarrow \sum_\mu M_\mu \rho_S M_\mu^\dagger ~,\\
&& M_\mu = {}_E\langle\mu | U_{SE} | 0\rangle_E ~,~~
   \sum_\mu M_\mu^\dagger M_\mu = I ~,\nonumber
\end{eqnarray}
using a complete basis for the environment $\{|\mu\rangle_E\}$. Since
the reduced density matrix has the same structure as the probabilistic
ensemble of classical statistical mechanics, it can be described in the
same language. But a ``quantum jump" mechanism is still needed to explain
how an entangled system-environment state collapses to an unentangled
system eigenstate in every single experimental run.

Generically the environment has a much larger number of degrees of freedom
than the system. Then, in the Markovian approximation which assumes that
information leaked from the system does not return, the evolution of the
reduced density matrix for the system can be expressed by the Lindblad
master equation \cite{Lindblad,GKS}:
\begin{eqnarray}
\label{mastereqn}
{d\over dt}\rho_S &=& i[\rho_S,H] + \sum_{\mu}{\cal L}[L_\mu]\rho_S ~,\\
{\cal L}[L_\mu]\rho_S &=& L_\mu\rho_S L_\mu^\dagger
  - {1\over2} \rho_S L_\mu^\dagger L_\mu - {1\over2} L_\mu^\dagger L_\mu\rho_S ~.
\label{Lindbladop}
\end{eqnarray}
The terms on the r.h.s. involving sum over $\mu$ modify the unitary
Schr\"odinger evolution, while $Tr(d\rho_S/dt)=0$ preserves the total
probability. When $H=0$, the fixed point of the evolution is a diagonal
$\rho_S$, in the basis that diagonalises $\{L_\mu\}$. This prefered basis
is determined by the system-environment interaction. (When there is no
diagonal basis for $\{L_\mu\}$, the evolution leads to equipartition,
i.e. $\rho_S \propto I$.) Furthermore, the off-diagonal elements of
$\rho_S$ decay, due to destructive interference among environmental
contributions with varying phases (arising from a large number of random
elastic scattering events), which is known as decoherence.

This modification of a quantum system's evolution, due to its interaction
with the unobserved environmental degrees of freedom, provides a proper
ensemble description, and a quantitative understanding of how the
off-diagonal elements of $\rho_S$ decay \cite{Zeh,Wiseman}. Still, the
``measurement problem" is not fully solved until we find the quantum
jump process that can predict outcomes of individual experimental runs,
and that forces us to go beyond the dynamics of Eq.(\ref{schrevol}).

\subsection{Continuous Stochastic Measurement}

The von Neumann interaction is usually taken to be the first step of the
measurement process. It is continuous and deterministic, and creates
perfectly entangled ``Schr\"odinger cat" states between the measurement
eigenstates of the system and the pointer states of the apparatus (which
is part of the system's environment):
\begin{equation}
\label{symstate}
|\psi\rangle_S = \sum_i c_i |i\rangle_S ~,~~
|\psi\rangle_S |0\rangle_E \longrightarrow
\sum_i c_i |i\rangle_S |\tilde{i}\rangle_E ~.
\end{equation}
To complete the measurement process, it needs to be supplemented by the
probabilistic quantum jump that selects a particular $|i\rangle$ and
collapses the reduced $\rho_S$ to a projection operator. Although the
physical mechanism behind quantum jump is not understood, it is common to
attribute quantum jump to interactions of a system with its surroundings.
With this postulate, we can define which system-environment interactions
are measurement interactions: {\it A measurement interaction is the one
in which the apparatus does not remain, for whatever reasons, in a
superposition of pointer states.}

The quantum jump can be realised via a continuous stochastic process,
while retaining the ensemble interpretation. The familiar method is to
add noise to a deterministic evolution, converting it into a Langevin
equation \cite{Pearle,Gisin,Diosi}. Such a realisation is strongly
constrained by the properties of quantum measurement. To ensure
repeatability of measurement outcomes, the measurement eigenstates
need to be fixed points of the evolution. The attraction towards the
eigenstates as well as the noise have to vanish at the fixed points,
which requires the evolution dynamics to be either nonlinear or non-unitary.
Furthermore, lack of simultaneity in special relativity must not conflict
with outcome probabilities in multipartite measurements. For instance, one
can consider pausing (or even abandoning) measurement part of the way along,
and that must not conflict with the consistency of the results. Since the
Born rule is fully consistent with special relativity, a solution is to
demand that the Born rule be satisfied at every instant of the measurement
process, when one averages over the noise. It is indeed remarkable that
such a continuous stochastic process for quantum measurement exists
\cite{Gisin}. It uses a precise combination of the attraction towards
the eigenstates and unbiased white noise to reproduce the Born rule.
(Some variations of the stochastic process away from this specific form have
been studied \cite{Pearle}, but they fail to satisfy all the constraints
\cite{PRLcomments}.)

The relation between the attraction towards the eigenstates and the noise,
needed to make the Born rule a constant of evolution in Ref.\cite{Gisin},
implies that the environmental degrees of freedom contribute to either
both the attraction and the noise (these degrees of freedom would be
considered the apparatus) or to neither of them (these degrees of freedom
can be ignored). This division strongly indicates a common origin for the
deterministic and the stochastic contributions to the measurement evolution,
quite reminiscent of the fluctuation-dissipation theorem of statistical
mechanics which is a consequence of both diffusion and viscous damping
arising from the same underlying molecular scattering. Such an intrinsic
property of quantum measurement dynamics would be an important clue to
figuring out what may lie beyond---an underlying theory of which quantum
measurement would be an effective process.

Understanding quantum measurement as a stochastic process is the focus of
our investigation, and we analyse its ingredients in detail in Section II.
With the technological progress in making quantum devices, such an analysis
is not just a formal theoretical curiosity, but is also a necessity for
increasing accuracy of quantum control and feedback \cite{Wiseman}.
A practical situation is that of the weak measurement \cite{Aharonov},
where information about the measured observable is extracted from the
system at a slow rate. Such a stretching out of the time scale would
allow one to monitor how the system state collapses to an eigenstate of
the measured observable, and to track properties of the intermediate
states created along the way by an incomplete measurement. Knowledge of
what really happens in a particular experimental run (and not the ensemble
average) would be invaluable in making quantum devices more efficient
and stable.

\subsection{Beyond Quantum Mechanics}

The stochastic measurement process does provide a continuous interpolation
of projective measurement. But, its nonlinear dynamics is distinct from the
unitary Schr\"odinger evolution, and one wonders how it may arise as an
effective description from a theory more fundamental than quantum mechanics.
Over the years, a variety of theoretical approaches have been proposed to
either solve or bypass the quantum measurement problem.

Some of the approaches that go beyond quantum mechanics are physical, e.g.
introduction of hidden variables with novel dynamics or ignored interactions
with known dynamics. Examples include Bohmian mechanics \cite{Bohm}, GRW
and CSL spontaneous collapse mechanisms \cite{Ghirardirev,Bassirev} and
modification of quantum rules due to gravity \cite{Penrose}. Some others
philosophically question what is real and what is observable, in principle
as well as by human beings with limited capacity. Examples include the
``many worlds" interpretation \cite{Everett} that assigns a distinct world
(i.e. an evolutionary branch) to each probabilistic outcome, and the
consistent histories formalism \cite{GMH}. Although these attempts are not
theoretically inconsistent, none of them have been positively verified by
experiments---only bounds exist on their parameters.

In this work, we reanalyse the quantum trajectory formalism for state
collapse (earlier reviews can be found in Refs.\cite{qtraj1,qtraj2}),
to achieve a deeper understanding of its dynamics. It is a particular
case of the class of stochastic collapse models that add a measurement
driving term and a random noise term to the Schr\"odinger evolution of
Eq.(\ref{schrevol}) \cite{Bassirev}. We treat these terms in an effective
theory approach, without assuming a specific collapse basis (e.g. energy
or position basis) or a specific collapse interaction (e.g. gravity or
some other universal interaction). This approach allows us to address the
possibility that the collapse process is non-universal, and the signal
amplification during the system-apparatus measurement interaction may be
responsible for it. The GRW/CSL models have not explored this possibility
much, and have typically focused on a particular collapse basis with a
particular collapse interaction.

Our approach is motivated by recent experimental advances in monitoring
quantum evolution during weak measurements on superconducting transmon
qubits \cite{Vijay,Murch}, where the collapse basis as well as the
system-apparatus interaction strength can be varied by changing the
circuit parameters and without changing the apparatus size. With suitable
choice of parameters, quantum trajectories interpolating from the initial
state to the final projected state have been observed \cite{Murch},
setting up a stage where the validity of the stochastic collapse paradigm
during measurement can be experimentally tested in detail. Such effective
theory tests would then impose restrictions on any extension of the
standard quantum theory, and that is what we aim for.

We formulate our model in the next Section, separating the quantum
collapse trajectories into a geodesic evolution part and a fluctuating
part on top of it. This separation allows us to point out that the
Born rule is equivalent to a fluctuation-dissipation type of relation
between the two parts, and we demonstrate that for two different types
of noise in Sections III and IV. We conclude with a discussion on the
implications of this property and the possible physical origin of the
noise.

\section{Quantum Geodesic Collapse}

In what follows, to keep the analysis simple, we concentrate only on the
evolution due to the effective system-apparatus interaction that leads to
measurement; other contributions to the system evolution can be added later
when needed. It is also convenient to abbreviate $\rho_S$ as $\rho$. We now
proceed to construct the complete quantum collapse dynamics in steps.

\subsection{A Single Geodesic Trajectory}

Assuming that the projective measurement results from a continuous geodesic
evolution of an initially pure quantum state to an eigenstate $|i\rangle$
of the measured observable, i.e. a great circle on the unit sphere in the
Hilbert space, one arrives at the nonlinear evolution equation \cite{dmqm}:
\begin{equation}
\label{projmeas}
{d\over dt}\rho = g\left[ \rho P_i + P_i\rho - 2\rho~Tr(P_i\rho) \right] ~.
\end{equation}
Here $t$ is the ``measurement time", the coupling $g$ represents the
strength of the system-apparatus interaction, while $gt$ is dimensionless.
This simple nonlinear evolution equation describing an individual quantum
trajectory has several noteworthy properties \cite{wavefnevol}.

\noindent(a)
In addition to maintaining $Tr(\rho)=1$, the evolution takes pure states
to pure states. $\rho^2=\rho$ implies
\begin{equation}
\label{evolpure}
{d\over dt}\big(\rho^2 - \rho\big)
= \rho{d\over dt}\rho + \big({d\over dt}\rho\big)\rho - {d\over dt}\rho = 0 ~.
\end{equation}
Thus the component of the state along $P_i$ grows at the expense of the other
orthogonal components.

\noindent(b)
Each projective measurement outcome is the fixed point of the
deterministic evolution:
\begin{equation}
{d\over dt}\rho = 0 \quad{\rm at}\quad
\rho_i^* = P_i\rho P_i/Tr(P_i\rho) ~.
\end{equation}
The fixed point nature of the evolution makes the measurement outcome
consistent on repetition.

\noindent(c)
In a bipartite setting (which includes the decoherence scenario),
the complete set of projection operators can be selected as
$\{P_i\} = \{P_{i_1} \otimes P_{i_2}\}$, with $\sum_i P_{i_k} = I_k$.
Since the evolution is linear in the projection operators, a sum over the
unobserved projection operators and a partial trace over the unobserved
degrees of freedom produces the same equation (and hence the same fixed
point) for the reduced density matrix for the system, as long as $g$ is
independent of the environment. This decoupling from the environment
forbids any possibility of superluminal signalling. Moreover, the evolution
purifies the state; for example, a qubit state in the interior of the Bloch
sphere evolves to a fixed point on its surface.

\noindent(d)
For pure states, the geodesic evolution equation is just
(using the notation of Eq.(\ref{Lindbladop}))
\begin{equation}
{d\over dt}\rho = -2g~{\cal L}[\rho] P_i ~.
\end{equation}
Compared to Eq.(\ref{mastereqn}), here the Lindblad operator acts on the
pointer state, the density matrix plays the role of $L_\mu$, and the sign
is reversed. This structure hints at an action-reaction relationship between
the processes of decoherence and collapse. Note here that both $\rho$ and
$P_i$ are projection operators, and after the von Neumann interaction
creates a symmetric entangled state of the system and the apparatus as in
Eq.(\ref{symstate}), it is a matter of subjective choice to consider whether
the system decoheres the apparatus or the apparatus decoheres the system.
In particular, $P_i$ can be looked upon as the apparatus state influenced
by the system operator $\rho$. Adding
$0 = {\cal L}[\rho]P_i - {\cal L}[\rho]P_i$
to the joint system-apparatus evolution equation, one can then envision
the following break-up during the measurement process: ${\cal L}[\rho]P_i$
combined with the apparatus dynamics decoheres the apparatus state $P_i$
(it cannot remain in superposition by definition), and the equal and
opposite $-{\cal L}[\rho]P_i$ combined with the system dynamics collapses
the system state $\rho$. Details of such a scenario remain to be worked out.

\noindent(e)
The limit $gt\rightarrow\infty$ corresponds to projective measurement,
while small $gt$ values describe weak measurements. Asymptotic convergence
to the fixed point is exponential,
$||\rho-P_i|| \sim e^{-2gt}$ as $t\rightarrow\infty$,
similar to the charging of a capacitor.

These properties make Eq.(\ref{projmeas}) a legitimate candidate for
describing the collapse of a quantum state during measurement, modeling
the single quantum trajectory specific to a particular experimental run.
It represents a superoperator that preserves Hermiticity, trace and
positivity, but is nonlinear.

\subsection{Ensemble of Geodesic Trajectories}

We next need a separate criterion for selection of $P_i$, to reproduce the
stochastic ensemble interpretation of quantum measurement. This choice of
``quantum jump" requires a particular $P_i$ to be picked with probability
$Tr(P_i\rho(t=0))$ as per the Born rule. Picking one of the $P_i$ at the
start of the measurement, and leaving it unaltered thereafter, is unsuitable
for gradual weak measurements, and we look for other ways to combine the
evolution trajectories for different $P_i$.

Let $w_i$ be the weight of the evolution trajectory for $P_i$, with
$\sum_i w_i = 1$. We want to find real $w_i(t)$, as some functions
of $\rho(t)$, that reproduce the well-established quantum behaviour.
The geodesic trajectory averaged evolution of the density matrix during
measurement is:
\begin{equation}
\label{trajavmeas}
{d\over dt}\rho = \sum_i w_i ~ g\left[\rho P_i + P_i\rho - 2\rho~Tr(P_i\rho)
                               \right] ~.
\end{equation}
Irrespective of the choice for $w_i$, this evolution maintains the properties
(a)-(d) described in the previous subsection, i.e. preservation of purity,
fixed point nature of all $P_i$, decoupling from environment, and a role
reversal relation with Lindblad operators.

With the decomposition, $\rho = \sum_{jk} P_j \rho P_k$, the projected
components of the density matrix evolve as
\begin{equation}
{d\over dt}\big(P_j \rho P_k\big)
= P_j \rho P_k ~g\Big[ w_j + w_k - 2\sum_i w_i~Tr(P_i\rho) \Big] .
\end{equation}
Independent of the choice of $\{w_i\}$, we have the identity,
\begin{eqnarray}
{2\over P_j \rho P_k}{d\over dt}\big(P_j \rho P_k\big)
&=& {1\over P_j \rho P_j}{d\over dt}\big(P_j \rho P_j\big) \nonumber\\
&+& {1\over P_k \rho P_k}{d\over dt}\big(P_k \rho P_k\big) ~,
\end{eqnarray}
with the consequence that the diagonal projections of $\rho$ completely
determine the evolution of all the off-diagonal projections. For an
$n$-dimensional quantum system, therefore, the evolution has only $n-1$
independent degrees of freedom. For one-dimensional projections,
$P_j \rho(t) P_j = d_j(t) P_j$ with $d_j\geq0$, we obtain:
\begin{equation}
\label{offdiagevol}
P_j \rho(t) P_k = P_j \rho(0) P_k
\left[ {d_j(t)~d_k(t) \over d_j(0)~d_k(0)} \right]^{1/2} ~.
\end{equation}
In particular, phases of the off-diagonal projections $P_j \rho P_k$ do not
evolve, in sharp contrast to what happens during decoherence. Also, their
asymptotic values, i.e. $P_j \rho(t\rightarrow\infty) P_k$, may not vanish,
whenever more than one diagonal $P_j \rho(t\rightarrow\infty) P_j$ remain
nonzero.

It is easily seen that when all the $w_i$ are equal, no information is
extracted from the system by the measurement and $\rho$ does not evolve.
More generally, the diagonal projections evolve according to:
\begin{equation}
\label{diagevol}
{d\over dt}d_j = 2g~d_j \big( w_j - \sum_i w_i d_i \big) .
\end{equation}
Here, with $\sum_i d_i=1$, $\sum_i w_i d_i \equiv w_{\rm av}$ is the
weighted average of $\{w_i\}$. Clearly, the diagonal projections with
$w_j > w_{\rm av}$ grow and the ones with $w_j < w_{\rm av}$ decay.
Any $d_j$ that is zero initially does not change, and the evolution is
therefore restricted to the subspace spanned by all the $P_j\rho(t=0)P_j
\ne 0$. These features are stable under small perturbations of the
density matrix.

A naive guess for the trajectory weights is the ``instantaneous Born rule",
i.e. $w_j = w_j^{IB} \equiv Tr(\rho(t)P_j)$ throughout the measurement
process. It avoids logical inconsistency in weak measurement scenarios,
where one starts the measurement, pauses somewhere along the way, and
then restarts the measurement. In this situation, the geodesic trajectory
averaged evolution is:
\begin{equation}
{d\over dt}\big(P_j \rho P_k\big)
= P_j \rho P_k ~ g\Big[ w_j^{IB} + w_k^{IB} - 2\sum_i (w_i^{IB})^2 \Big] .
\end{equation}
This evolution converges towards the subspace specified by the largest
diagonal projections of the initial $\rho(t=0)$, i.e. the closest fixed
points. It is deterministic too, and differs from Eq.(\ref{Bornrule}).
So $w_j=w_j^{IB}$ is unphysical, and we need to find $w_i$ with stochastic
behaviour that would reproduce the Born rule.

\subsection{Addition of Noise}

Instead of heading towards the nearest fixed point, quantum trajectories
can be made to wander around and explore other possibilities by adding
noise to their dynamics. The combination of geodesics and fluctuations
generically appears in variational calculus, easily seen in the path
integral framework for instance. Noisy fluctuations are also expected
to contribute to the measurement process \cite{Korotkov,Vijay,Murch}.
So we search for a suitable noise, which when combined with the geodesics
already described would reproduce Eq.(\ref{Bornrule}). The existence of
such a noise is a hypothesis, to be verified by its explicit construction
and evaluation of its consequences. In order to not lose the handsome
features of the geodesic trajectories, we make the noise part of the
trajectory weights $w_i$, while retaining $\sum_i w_i=1$. In describing
quantum measurement as a stochastic process, two commonly considered
situations are ``white noise" and ``shot noise", with the corresponding
evolution dynamics labeled ``quantum diffusion" and ``binary quantum jump"
respectively \cite{qtraj1,qtraj2}, and we analyse them in turn in the
next two Sections. It should be noted that our formalism allows us to
freely vary the size of the noise, unlike the fixed specific values
considered in earlier works, and explore the consequences.

\section{Quantum Diffusion}

In the quantum diffusion model, unbiased and uncorrelated noise (i.e.
white noise) is added to the geodesic evolution. With a gradual addition
of the noise, the quantum trajectories remain continuous but become
non-differentiable. The deterministic evolution equation in the Hilbert
space gets converted to a stochastic Langevin type equation, and we need
to find the magnitude of the frequency independent noise that makes the
measurement process consistent with the Born rule.

\subsection{Constraint on White Noise}

Results of the previous Section take a considerably simpler form in
case of the smallest quantum system, i.e. the two-dimensional qubit with
$|0\rangle$ and $|1\rangle$ as the measurement eigenstates. Evolution of
the density matrix during the measurement, Eqs.(\ref{diagevol}) and
(\ref{offdiagevol}), is then given by:
\begin{eqnarray}
\label{Strat1qd}
{d\over dt}\rho_{00} &=& 2g(w_0-w_1)~\rho_{00}~\rho_{11} ~, \\
\rho_{01}(t) &=& \rho_{01}(0) \left[ {\rho_{00}(t)~\rho_{11}(t)
               \over \rho_{00}(0)~\rho_{11}(0)} \right]^{1/2} ~,
\label{Strat1qod}
\end{eqnarray}
Because of $\rho_{11}(t)=1-\rho_{00}(t)$ and $w_1(t)=1-w_0(t)$, only one
independent variable describes the evolution of the system. Selecting the
trajectory weights as addition of real white noise to the ``instantaneous
Born rule", we have
\begin{equation}
\label{oneqweight}
w_0 - w_1 = \rho_{00} - \rho_{11} + \sqrt{S_\xi}~\xi ~.
\end{equation}
Here, $\llangle\xi(t)\rrangle=0$ is unbiased,
$\llangle\xi(t)\xi(t')\rrangle = \delta(t-t')$
fixes the normalisation of $\xi(t)$, and $S_\xi$ is the spectral density
of the noise.

\begin{figure}[b]
\epsfxsize=8.6truecm
\centerline{\epsfbox{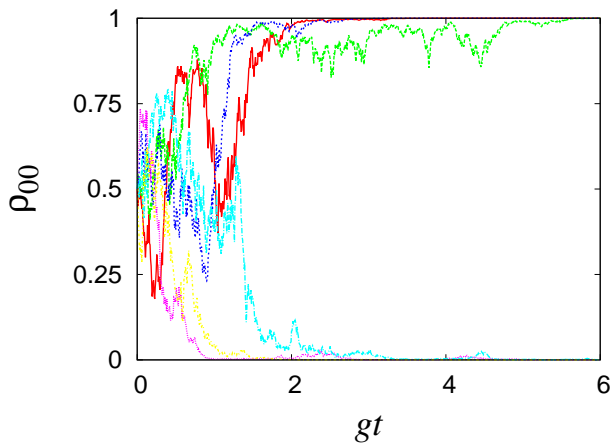}}
\caption{Individual quantum evolution trajectories for the initial state
$\rho_{00}=0.5$, the measurement eigenstates $\rho_{00}=0,1$, and in
presence of measurement noise satisfying $gS_\xi=1$.}
\label{singletraj}
\end{figure}

Equations (\ref{Strat1qd},\ref{oneqweight}) define a stochastic differential
process on the interval $[0,1]$. The fixed points at $\rho_{00}=0,1$
are perfectly absorbing boundaries where the evolution stops. In general,
a quantum trajectory would zig-zag through the interval before ending at
one of the two boundary points. Some examples of such trajectories are
shown in Fig.\ref{singletraj}.

Let $P(x)$ be the probability that the initial state with $\rho_{00}=x$
evolves to the fixed point at $\rho_{00}=1$. Obviously, $P(0)=0$,
$P(0.5)=0.5$, $P(1)=1$. Two extreme situations are easy to figure out.
When there is no noise, the evolution is governed by the sign of
$\rho_{00}-\rho_{11}$ and the trajectory monotonically approaches the
fixed point closest to the starting point.
\begin{equation}
S_\xi=0: \quad P(x) = \theta(x-0.5) ~.
\end{equation}
Also, when $\rho_{00}-\rho_{11}$ is negligible compared to the noise,
symmetry of the evolution makes both eigenstates equiprobable, i.e.
$P(x)=0.5$ for $S_\xi\rightarrow\infty$.

The stochastic evolution equations, Eqs.(\ref{Strat1qd},\ref{oneqweight}),
are in the Stratonovich form. For further insight into the evolution, and
for numerical simulations, it is instructive to convert them into the It\^o
form that specifies forward evolutionary increments \cite{StrIto}:
\begin{eqnarray}
\label{Ito1qd}
d\rho_{00} &=& 2g~\rho_{00}~\rho_{11}
               \big(\rho_{00}-\rho_{11}\big) (1-gS_\xi)~dt \nonumber\\
           &+& 2g\sqrt{S_\xi}~\rho_{00}~\rho_{11}~dW ~.
\end{eqnarray}
Here the stochastic Wiener increment $dW=\xi dt$ obeys
$\llangle dW(t)\rrangle=0$, $\llangle (dW(t))^2\rrangle=dt$, and can be
modeled as a random walk. The first term on the r.h.s. produces drift in
the evolution, while the second term gives rise to diffusion.

The evolution with no drift, i.e. the pure Wiener process, is particularly
interesting. In that case, after averaging over the stochastic noise, the
Born rule is a constant of evolution \cite{Pearle,Gisin}:
\begin{equation}
\label{BRdiff}
\llangle d\rho_{00} \rrangle=0 ~\Longleftrightarrow~ gS_\xi=1 ~.
\end{equation}
More explicitly, starting at $x$, one moves forward to $x+\epsilon$ with
some probability, moves backward to $x-\epsilon$ with the same probability,
and stays put otherwise. On balancing the probabilities,
$P(x) = \alpha (P(x+\epsilon)+P(x-\epsilon)) + (1-2\alpha)P(x)$,
and we get
\begin{equation}
gS_\xi=1: \quad P(x+\epsilon) - 2P(x) + P(x-\epsilon) = 0 ~.
\end{equation}
The general solution, independent of the choice of $\epsilon$, is that
$P(x)$ is a linear function of $x$. Imposing the boundary conditions,
$P(0)=0$ and $P(1)=1$, we obtain $P(x)=x$, which is the Born rule.
Note that specific choices of $g$, $\epsilon$ and $\alpha$ only alter
the rate of evolution, but not this final outcome.

\begin{figure}[b]
\epsfxsize=8.6truecm
\centerline{\epsfbox{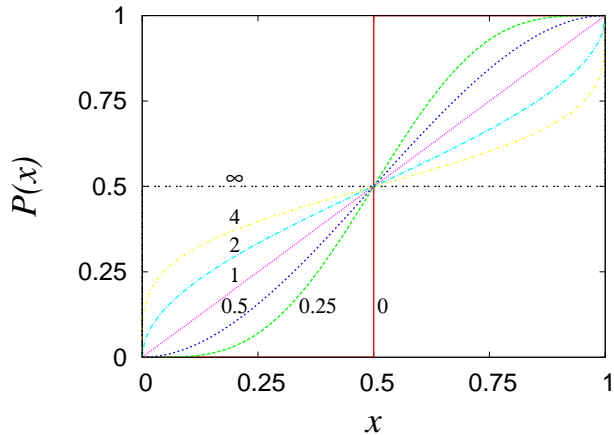}}
\caption{Probability that the initial qubit state $\rho_{00}=x$ evolves
to the measurement eigenstate $\rho_{00}=1$, for different magnitudes
of the measurement noise. The $gS_\xi$ values label the curves.}
\label{pxvsx}
\end{figure}

Going further, we performed numerical simulations of the stochastic
evolution for several values of $gS_\xi$, and the results are presented
in Fig.\ref{pxvsx}. We used the integrated form of Eq.(\ref{Strat1qd})
over a short time step $g\Delta t \ll 1$:
\begin{eqnarray}
&& {\rho_{00}(t+\Delta t) \over \rho_{11}(t+\Delta t)} = 
   {\rho_{00}(t) \over \rho_{11}(t)} ~ e^{2g\Delta t\overline{w}} ~, \\
&& \overline{w} = {1\over\Delta t} \int_t^{t+\Delta t} (w_0-w_1) ~ dt ~.
\end{eqnarray}
$\overline{w}$ was generated as a Gaussian random number with mean
$\rho_{00}(t)-\rho_{11}(t)$ and variance $S_\xi/\Delta t$. We averaged
the results over a million trajectories at each simulation point.
The data clearly show the cross-over from evolution with no noise to
evolution with only noise, and the Born rule behaviour appears for
$gS_\xi=1$.

We point out that with $gS_\xi=1$, Eq.(\ref{Ito1qd}) is the same as the
corresponding result of Ref.\cite{Gisin}. But our strategy of breaking up
the evolution into geodesic and fluctuating parts allows us to analyse the
two contributions separately, e.g. in the fluctuation-dissipation relation
described later in Section III.C, and explore the implications. Also, we
can easily extend the result to $n$-dimensional orthogonal measurements
as in Eq.(\ref{multidimweight}).

The preceding results are valid for binary orthogonal measurements on any
quantum system, with the replacement $\rho_{ii} \rightarrow Tr(\rho P_i)$.
One way to extend them to a larger set of $P_i$, is to express non-binary
orthogonal projection operators as a product of mutually commuting binary
projection operators, and then treat each binary projection as per the
preceding analysis with its own stochastic noise \cite{Gisin}. An alternate
way to implement $n$-dimensional orthogonal quantum measurements is to
observe that $P_0-P_1$ is one of the Cartan generators of $SU(n)$, and
it can be rotated to any of the other Cartan generators of $SU(n)$ by the
unitary symmetry. Such a rotation of Eq.(\ref{oneqweight}) allows us to
fix all the orthonormal set of weights as ($k=1,2,\ldots,n-1$):
\begin{equation}
\label{multidimweight}
\sum_{i=0}^{k-1} w_i - kw_k = \sum_{i=0}^{k-1} \rho_{ii} - k\rho_{kk}
                            + \sqrt{k(k+1)S_\xi \over 2} ~ \xi_k ~.
\end{equation}
Here, $\xi_k(t)$ are independent white noise terms. The condition for the
evolution to be a pure Wiener process, and consequently satisfy the Born
rule, remains $gS_\xi=1$. With this condition, the evolution equation in
the Stratonovich form is Eq.(\ref{diagevol}), while in the It\^o form it
is given by
\begin{eqnarray}
&& d(d_j) = 2\sqrt{g}~d_j \big( \tilde{w}_j - \sum_i \tilde{w}_i d_i \big) , \\
&& \sum_{i=0}^{k-1} \tilde{w}_i - k\tilde{w}_k = \sqrt{k(k+1) \over 2} ~ dW_k ~.
\end{eqnarray}

Two important properties, arising from the parametric freedom of the
stochastic evolution analysed here, are worth stressing:\\
(a) The equations need specification of only the first two moments of
the stochastic noise, whether $\xi(t)$ or $dW(t)$. We can use the remaining
freedom in the specification of the noise to simplify our analysis as much
as possible. The binary or $Z_2$ noise is the simplest choice, producing
two stochastic possibilities at every time step.\\
(b) The solution to the equations implies that a complete measurement
formally takes infinite time, and that may not be a desirable feature
\cite{PRLcomments}. The formal ``measurement duration" can be made finite
by making the coupling $g$ time-dependent. Such a stochastic differential
process does not have a time translation symmetry, but the change of
variables does not alter the measurement outcomes because the Born rule is
satisfied at every instant during the measurement process. With such a
modification, the only change required in the results described here is
to replace $gt$ by $\tau\equiv\int_0^t g(t') dt'$. The choice of $g(t)$
is not unique, and its physical meaning would depend on the nature of
the system-apparatus interaction. For example, with $g(t)=1/(1-t^2)$,
the measurement interval becomes $t\in[0,1]$ and Eq.(\ref{Strat1qd})
can be written in coupling-free form as
\begin{equation}
{d\over ds}\rho_{00} = 2(\rho_{00}-\rho_{11}+\tilde{\xi})
                       ~ \rho_{00} ~ \rho_{11} ,
\end{equation}
where $s=\tanh^{-1}t\in[0,\infty]$ and
$\llangle\tilde{\xi}(s)\tilde{\xi}(s')\rrangle = \delta(s-s')$.

\subsection{Born Rule Satisfying Trajectory Ensemble}

Henceforth, we impose $gS_\xi=1$, and focus on the set of the Born rule
satisfying quantum trajectories. For a qubit, the map
\begin{equation}
\tanh(z) = 2\rho_{00}-1 = \rho_{00}-\rho_{11} ~,
\end{equation}
between $\rho_{00}\in[0,1]$ and $z\in[-\infty,\infty]$, simplifies the
evolution equations. Both the Stratonovich and the It\^o forms,
Eqs.(\ref{Strat1qd},\ref{Ito1qd}), then have the same structure:
\begin{eqnarray}
{dz\over dt} &=& g\tanh(z) + \sqrt{g}~\xi ~, \\
dz &=& g\tanh(z)~dt + \sqrt{g}~dW ~.
\label{zevol1q}
\end{eqnarray}
In terms of $z(t)$, the density matrix has the form
\begin{eqnarray}
\label{denmatz}
\rho(z(t)) &=& \pmatrix{ {1+\tanh(z(t))\over2}
               & {\rho_{01}(z(0))~{\rm sech}(z(t)) \over {\rm sech}(z(0))} \cr
                 {\rho_{10}(z(0))~{\rm sech}(z(t)) \over {\rm sech}(z(0))}
               & {1-\tanh(z(t))\over2} \cr} \nonumber\\
           &=& {\rm sech}(z(t)) \pmatrix{ {1\over2} e^{z(t)}
               & {\rho_{01}(z(0)) \over {\rm sech}(z(0))} \cr
                 {\rho_{10}(z(0)) \over {\rm sech}(z(0))}
               & {1\over2} e^{-z(t)} \cr} ,
\end{eqnarray}
and average over the stochastic noise provides the Born rule constraint
$\llangle\tanh(z(t))\rrangle = \tanh(z(0))$.

\begin{figure}[b]
\epsfxsize=8.6truecm
\centerline{\epsfbox{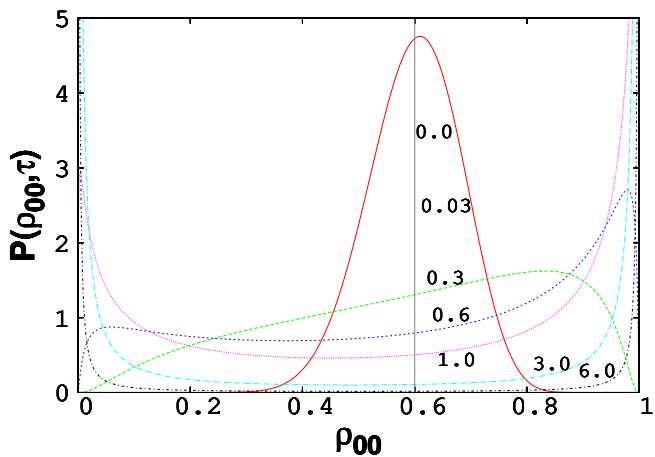}}
\caption{
Evolving distribution of the quantum measurement trajectories for a qubit
state initialised to $\rho_{00}=\delta(0.6)$. The curves are labeled by
values of the evolution parameter $\tau \equiv \int_0^t g~dt$, and they
accumulate on the eigenstates as $\tau\rightarrow\infty$.}
\label{distevol}
\end{figure}

The stochastic Langevin evolution can be converted to the Fokker-Planck
equation to obtain the collective behaviour of the quantum trajectories.
For measurement of a single qubit with $gS_\xi=1$, the probability
distribution of trajectories, $p(\rho_{00},t)$ or $p(z,t)$, satisfies:
\begin{eqnarray}
{\partial p(\rho_{00},t) \over \partial t} &=&
  2g {\partial^2 \over \partial^2 \rho_{00}}
  \left( \rho_{00}^2 (1-\rho_{00})^2 p(\rho_{00},t) \right) ~, \\
{\partial p(z,t) \over \partial t} &=&
  - g{\partial \over \partial z}(\tanh(z) p(z,t))
  + {g\over2} {\partial^2 \over \partial z^2} p(z,t) ~. \nonumber
\end{eqnarray}
With the initial condition $p(\rho_{00},0)=\delta(x)$, this equation
can be solved exactly \cite{Pearle,Gisin}. The solution consists of two
non-interfering peaks with areas $x$ and $1-x$, monotonically traveling
to the boundaries at $\rho_{00}=1$ and $0$ respectively. In terms of the
variable $z$, the two peaks are diffusing Gaussians, with centres at
$z_\pm(t) = \tanh^{-1}(2x-1) \pm gt$ and common variance $gt$,
\begin{eqnarray}
\label{qdtrajdist}
p(z,t) = {1\over\sqrt{2\pi gt}}
       &\Big(& x \exp\big[-{(z-z_+)^2 \over 2gt}\big] \\
       &+&  (1-x) \exp\big[-{(z-z_-)^2 \over 2gt}\big] \Big) ~. \nonumber
\end{eqnarray}
The two peaks reach the boundaries only asymptotically:
\begin{equation}
p(\rho_{00},\infty) = x~\delta(\rho_{00}-1) + (1-x)~\delta(\rho_{00}) ~.
\end{equation}
A particular case of how a narrow initial distribution splits into two
components that evolve to the measurement eigenstates is illustrated in
Fig.\ref{distevol}. For $gt>10$, 99\% of the probability is within 1\%
of the two fixed points. Subsequent convergence to projective measurement
is exponential, e.g. 99.9\% of the probability is within 0.1\% of the two
fixed points for $gt>15$.

Upon taking the ensemble average,
\begin{eqnarray}
\int_\infty^\infty \tanh(z(t)) ~p(z,t)~ dz = 2x-1 ~, \\
\int_\infty^\infty {\rm sech}(z(t)) ~p(z,t)~ dz = e^{-gt/2} {\rm sech}(z(0)) ~.
\end{eqnarray}
The resultant expectation value of the density matrix is
(cf. Eq.(\ref{denmatz})),
\begin{equation}
\label{onequbitevol}
\llangle\rho(t)\rrangle = \pmatrix{ x & e^{-gt/2} \rho_{01}(0)   \cr
                                    e^{-gt/2} \rho_{10}(0) & 1-x \cr} ,
\end{equation}
where the diagonal elements do not evolve and the off-diagonal elements
decay exponentially. Directly, the constraint of Eq.(\ref{Strat1qod}),
and Eq.(\ref{Ito1qd}), also give:
\begin{eqnarray}
d\rho_{01} &=& \rho_{01} \left( 1
            + {\rho_{11}-\rho_{00} \over 2\rho_{00}\rho_{11}} d\rho_{00}
            - {1 \over 8\rho_{00}^2\rho_{11}^2} d\rho_{00}^2 \right) \nonumber\\
&\Longrightarrow& \llangle d\rho_{01} \rrangle = \rho_{01} (-g~dt/2) ~.
\end{eqnarray}
We observe that this mixed state results from averaging individual pure
state fluctuating trajectories. Note that in the conventional ensemble
interpretation (cf. Eq.(\ref{Bornrule})), all the expectation values are
linear functions of the density matrix and so depend only on
$\llangle\rho(t)\rrangle$.

The Lindblad master equation for the same system gives evolution identical
to Eq.(\ref{onequbitevol}), with the single decoherence operator
$L_\mu = \sqrt{\gamma}\sigma_3$ and $\gamma=g/4$:
\begin{eqnarray}
{d\over dt}\rho &=& \gamma(\sigma_3\rho\sigma_3 - \rho) ~, \\
\rho(t) &=& {1+e^{-2\gamma t} \over 2}\rho(0) 
         +  {1-e^{-2\gamma t} \over 2}\sigma_3\rho(0)\sigma_3 ~.
\end{eqnarray}
Amazingly, the nonlinear stochastic evolution of the density matrix,
after averaging over the noise, becomes linear evolution described by
a completely positive trace-preserving map.

The result can also be expressed in the Kraus decomposed orthogonal form as:
\begin{eqnarray}
\rho(t) = M_0 \rho(0) M_0 + M_3 \rho(0) M_3 ~, \\
M_0^2 + M_3^2 = I ~,~~ Tr(M_0 M_3) =0 ~,
\end{eqnarray}
where, with $\cosh\epsilon = e^{2\gamma t} = e^{gt/2}$,
\begin{eqnarray}
M_0 &=& \sqrt{e^{-\gamma t}\cosh(\gamma t)} I 
     =  {\cosh(\epsilon/2)\over\sqrt{\cosh\epsilon}} I ~, \\
M_3 &=& \sqrt{e^{-\gamma t}\sinh(\gamma t)} \sigma_3
     =  {\sinh(\epsilon/2)\over\sqrt{\cosh\epsilon}} \sigma_3 ~.
\end{eqnarray}
The Kraus decomposition can also be performed in a symmetric but
non-orthogonal form as:
\begin{eqnarray}
&& \rho(t) = M_+ \rho(0) M_+ + M_- \rho(0) M_- ~, \\
&& M_+^2 + M_-^2 = I ~,~~ Tr(M_+^2) = Tr(M_-^2) ~,
\end{eqnarray}
\begin{equation}
M_\pm = {e^{\pm\epsilon/2} \over \sqrt{2\cosh\epsilon}}
        \left( {I+\sigma_3 \over 2} \right)
      + {e^{\mp\epsilon/2} \over \sqrt{2\cosh\epsilon}}
        \left( {I-\sigma_3 \over 2} \right) .
\end{equation}
This is the form used to describe binary weak measurement evolution in
Ref.\cite{oreshkov}, with $\epsilon$ as the evolution parameter. Then
the two evolution possibilities can be expressed as a biased walk,
\begin{eqnarray}
\label{binevol}
\rho(z,\epsilon) &=& M_+ \rho(z) M_+ + M_- \rho(z) M_- \nonumber\\
                 &=& p_+ \rho(z+\epsilon) + p_- \rho(z-\epsilon) ~,
\end{eqnarray}
with the parametrisation of Eq.(\ref{denmatz}), and
\begin{equation}
p_\pm = {\rm Tr}(M_\pm \rho M_\pm) = (1\pm\tanh(z)\tanh(\epsilon))/2 ~.
\end{equation}
Note that when $\rho(z)$ is a pure state, so are $\rho(z\pm\epsilon)$.
So the two contributions on r.h.s. of Eq.(\ref{binevol}) can be
considered two possible trajectories with unequal weights $p_\pm$;
it is indeed the finite duration integral of Eq.(\ref{zevol1q})
with $Z_2$ noise.


\subsection{Salient Features}

The evolution constraint that produces the Born rule, $gS_\xi=1$, relates
the strength of the geodesic evolution $g$ to the magnitude of the noise
$S_\xi$. So it is sensible to express it as a fluctuation-dissipation
relation.

For the white noise measurement, the geodesic parameter is
$\rho_{00}-\rho_{11}$. From Eq.(\ref{Ito1qd}), the size of
the fluctuations is, dropping the subleading $o(dt)$ terms,
\begin{equation}
\llangle (d\rho_{00}-d\rho_{11})^2 \rrangle =
  16g^2 S_\xi ~ \rho_{00}^2 \rho_{11}^2 ~ dt ~.
\end{equation}
The geodesic evolution term is, from Eq.(\ref{Strat1qd}) with $w_j$
replaced by its average $w_j^{IB}$,
\begin{equation}
(d\rho_{00}-d\rho_{11})_{\rm geo}
  = 4g(\rho_{00}-\rho_{11})\rho_{00}\rho_{11} ~ dt ~.
\end{equation}
Hence $gS_\xi=1$ amounts to the coupling-free relation:
\begin{equation}
\label{fdreldiff}
\llangle (d\rho_{00}-d\rho_{11})^2 \rrangle
  = 4\rho_{00}\rho_{11} {(d\rho_{00}-d\rho_{11})_{\rm geo}
                         \over \rho_{00}-\rho_{11}} ~.
\end{equation}
The proportionality factor between the noise and the damping term is not
a constant, because of the nonlinearity of the evolution, but it becomes
independent of $(g~dt)$ when the Born rule is satisfied.

In addition, our analysis has revealed the following notable aspects of
the quantum diffusion model:

\noindent
(1) Individual quantum trajectories maintain purity, even in the presence
of noise. Mixed states arise only when multiple quantum trajectories with
different evolutionary weights are combined.

\noindent
(2) Although the trajectory weights $w_i(t)$ are real and add up to one,
they are not restricted to the interval $[0,1]$, and so cannot be
interpreted as probabilities.

\noindent
(3) The measurement outcomes are independent of $\rho_{i\ne j}$, and so are
not affected by decoherence. In general, a different noise can be added to
the phases of $\rho_{i\ne j}$, without spoiling the described evolution of
$\rho_{ii}$. The Born rule imposes no constraint on that off-diagonal noise.
Measurement and decoherence can therefore be looked upon as independent and
complementary processes.

\noindent
(4) When the Born rule is satisfied, the measurement dynamics allows free
reparametrisation of the ``measurement time" but no other freedom. The
choice of measurement time is local between the system and the apparatus;
different interacting system-apparatus pairs can have different couplings
governing their collapse time scales.

\noindent
(5) The quantum trajectory distribution, given by Eq.(\ref{qdtrajdist})
and illustrated in Fig.\ref{distevol}, is fully determined in terms
of the single evolution parameter $\tau\equiv\int_0^t g(t')dt'$. In weak
measurement experiments on superconducting qubits \cite{Vijay,Murch},
the coupling $g$ is a tunable parameter and $\tau$ can be gradually
varied, e.g. in the range $[0,10]$. Such experiments can observe quantum
trajectory distributions in detail, and so can verify the theoretical
predictions.

\section{Binary Quantum Jump}

In the quantum jump model, a large but infrequent noise (i.e. shot noise)
is added to the geodesic evolution. The quantum trajectories are smooth
most of the time, except for the instances where sudden addition of the
noise makes them discontinuous. The measurement is often a binary process
in the Fock space, and sudden jumps terminate it, e.g. by emission of a
photon. Again, we need to find the magnitude of the noise that makes the
measurement process consistent with the Born rule.

\subsection{Constraint on Shot Noise}

Consider the binary measurement scenario, where the eigenstate $P_0$
is reached by continuous geodesic evolution, while the eigenstate $P_1$
is reached by a sudden jump. Then the density matrix evolution during
measurement is specified, with trajectory weights $w_i=\delta_{i0}$ and
binary shot noise $dN \in \{0,1\}$, as
\begin{equation}
\label{shotevol}
d\rho = g[\rho P_0 + P_0 \rho - 2\rho Tr(P_0 \rho)] dt + (P_1 - \rho) dN ~.
\end{equation}
The shot noise contribution is not infinitesimal; the density matrix
instantaneously jumps to $P_1$, whenever $dN=1$. Of course, the
probability of occurrence of $dN=1$ is an infinitesimal function of
$\rho$, and it has to vanish at the measurement eigenstate $\rho=P_0$.

For a single qubit, Eq.(\ref{shotevol}) reduces to:
\begin{eqnarray}
\label{shotevold}
d\rho_{00} &=& 2g~\rho_{00} \rho_{11}~dt - \rho_{00}~dN ~,\\
d\rho_{01} &=& g~\rho_{01} (\rho_{11}-\rho_{00})~dt - \rho_{01}~dN ~.
\label{shotevolod}
\end{eqnarray}
Once again, evolution of the off-diagonal elements is completely determined
in terms of that for the diagonal elements. The Born rule can be implemented
as a constant of evolution, constraining how often the jumps occur:
\begin{equation}
\label{BRshot}
\llangle d\rho_{00} \rrangle = 0 ~\Longleftrightarrow~
\llangle dN \rrangle = 2g~\rho_{11}~dt ~.
\end{equation}

From these evolution equations, an ensemble of quantum trajectories
can be constructed, allowing for two possibilities for $dN$ at every
instance. The $dN=0$ branch gradually keeps moving towards $\rho_{00}=1$
as a function of time, while the $dN=1$ branch stops evolving immediately
after the jump to $\rho_{00}=0$.

\subsection{Born Rule Satisfying Trajectory Ensemble}

Even though Eqs.(\ref{shotevold},\ref{shotevolod}) are not differential
equations in the usual sense, due to finite $dN$, they can be solved
exactly as a biased random walk process.

\begin{figure}[b]
\epsfxsize=8.6truecm
\centerline{\epsfbox{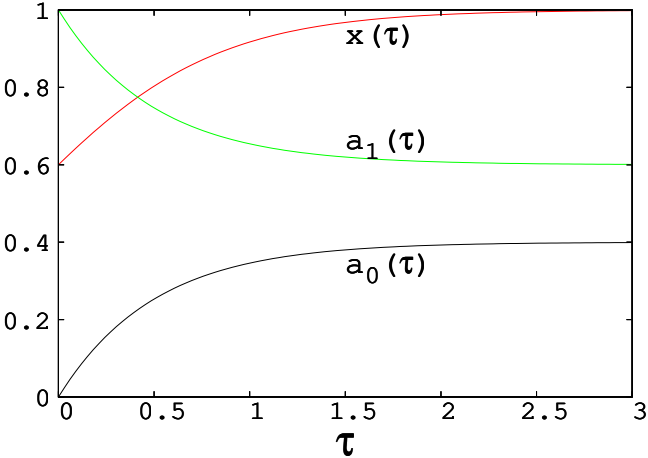}}
\caption{
Properties of the quantum measurement trajectories for quantum jump
evolution of a qubit. The initial state is $\rho_{00}(\tau=0)=0.6$,
and the evolution parameter is $\tau \equiv \int_0^t g(t') dt'$.
The initial distribution splits into a monotonically moving component
$a_1(\tau)~\delta(x(\tau))$ and a stationary component $a_0(\tau)~\delta(0)$,
which respectively move to the measurement eigenstates $\rho_{00}=1$ and
$\rho_{00}=0$ as $\tau\rightarrow\infty$.}
\label{jumpevol}
\end{figure}

Let the initial condition be $p(\rho_{00},0) = \delta(x)$. Because the
$dN=1$ evolution branch terminates at $\rho_{00}=0$, the solution consists
of two $\delta$-functions at any instant. The $\delta$-function at
$\rho_{00}=0$ steadily grows in size, while the $\delta$-function slowly
moving to $\rho_{00}=1$ gradually reduces in size. Explicitly,
\begin{eqnarray}
\label{qjtrajdist}
p(\rho_{00},t) &=& (x+(1-x)e^{-2gt})
                 ~ \delta\left(\frac{x}{x+(1-x)e^{-2gt}}\right) \nonumber\\
               &+& (1-x)(1-e^{-2gt}) ~ \delta(0) ~.
\end{eqnarray}
A particular case of how the variables in this distribution evolve is shown
in Fig.\ref{jumpevol}.

The corresponding distribution for the off-diagonal element also consists
of two $\delta$-functions, given by
\begin{eqnarray}
p(\rho_{01},t) &=& (x+(1-x)e^{-2gt})
                 ~ \delta\left(\frac{\rho_{01}(0)}{xe^{gt}+(1-x)e^{-gt}}\right)
                   \nonumber\\
               &+& (1-x)(1-e^{-2gt}) ~ \delta(0) ~.
\end{eqnarray}
Upon taking the ensemble average, the expectation value of the density
matrix becomes
\begin{equation}
\llangle\rho(t)\rrangle
  = \pmatrix{\rho_{00}(0) & e^{-gt} \rho_{01}(0) \cr
             e^{-gt} \rho_{10}(0) & \rho_{11}(0) \cr} .
\end{equation}
The exponential decay of the off-diagonal elements can also be obtained from
Eq.(\ref{shotevolod}) as:
\begin{equation}
\llangle d\rho_{01} \rrangle = \rho_{01} (-g~dt) ~.
\end{equation}

This result is again the solution of the Lindblad master equation for the same
system, with the single decoherence operator $L_\mu = \sqrt{\gamma}(P_0-P_1)$,
$\gamma=g/2$. The nonlinear evolution of quantum trajectories, after
averaging over their distribution, produces a linear completely positive
trace-preserving evolution for the density matrix. It can be expressed in
the Kraus decomposed form the same way as in the case of quantum diffusion.

\subsection{Salient Features}

The evolution constraint yielding the Born rule,
$\llangle dN \rrangle = 2g\rho_{11}dt$, relates the strength of the geodesic
evolution $g$ to the frequency of the noise $dN$. It can again be expressed
as a fluctuation-dissipation relation.

For the shot noise measurement, the geodesic parameter is $\rho_{00}$ and
$(dN)^2=dN$. Dropping the subleading $o(dt)$ terms, Eq.(\ref{shotevold})
gives the size of the fluctuations as
\begin{equation}
\llangle (d\rho_{00})^2 \rrangle = \rho_{00}^2 \llangle dN \rrangle ~.
\end{equation}
The geodesic evolution term, also from Eq.(\ref{shotevold}), is
\begin{equation}
(d\rho_{00})_{\rm geo} = 2g\rho_{00}\rho_{11} ~ dt ~.
\end{equation}
Hence, $\llangle dN \rrangle = 2g\rho_{11}dt$ amounts to the coupling-free
relation:
\begin{equation}
\label{fdrelshot}
\llangle (d\rho_{00})^2 \rrangle
  = \rho_{00}^2 {(d\rho_{00})_{\rm geo} \over \rho_{00}} ~.
\end{equation}
Once more, the noise is proportional to the damping term. Although the
proportionality factor differs from that in Eq.(\ref{fdreldiff}),
because of a different nonlinear evolution, it still becomes independent
of $(g~dt)$ when the Born rule is satisfied.

In addition, our analysis has brought out the following features of the
binary quantum jump model:

\noindent
(1) In the presence of shot noise, the quantum trajectories are monotonic,
and smooth except for infrequent discontinuous jumps. They still maintain
purity, and mixed states arise when multiple quantum trajectories with
different noise histories are combined.

\noindent
(2) The trajectory weights can be interpreted as probabilities, since the
shot noise has a direct probabilistic interpretation as a Poisson process.

\noindent
(3) Evolution of the diagonal $\rho_{ii}$ is independent of the off-diagonal
$\rho_{i\ne j}$, and so is unaffected by decoherence. So as in case of
quantum diffusion, measurement and decoherence can be looked upon as
independent and complementary processes. Also, free reparametrisation of
the ``measurement time" is allowed, when the Born rule is satisfied.

\noindent
(4) The measurement dynamics is local between the system and the apparatus.
The quantum trajectory distribution, given by Eq.(\ref{qjtrajdist}), is
fully determined in terms of the evolution parameter $\int_0^t g(t)dt$.
Weak measurement experiments in quantum optics should be able to verify
this theoretical prediction.

\section{Discussion}

We have described a quantum trajectory formalism for state collapse during
measurement, which replaces the discontinuous projective measurement by a
continuous stochastic process and remains consistent with the Born rule.
It supplements the Schr\"odinger evolution by addition of quadratically
nonlinear measurement terms:
\begin{eqnarray}
\label{eqncollapse}
d\rho &=& i[\rho,H]~dt \\
      &+& \sum_i w_i~g[ \rho P_i + P_i \rho - 2\rho~Tr(\rho P_i) ]~dt
       +  {\rm noise} ~. \nonumber
\end{eqnarray}
Instead of attributing the additional terms to novel interactions beyond the
standard quantum theory, we look at them as an effective description of the
system-apparatus measurement interaction that replaces the von Neumann
projection axiom. The task is then to figure out what restrictions such an
effective description imposes on the underlying unknown measurement dynamics
(including the type of noise that may be present), and whether or not the
necessary ingredients exist in the physical world. Nonlinear superoperator
evolution for the density matrix is avoided in quantum mechanics, because it
conflicts with the probability interpretation for mixtures of density matrices.
Nevertheless, nonlinear quantum evolutions need not be unphysical, and our
analysis in Section II shows that Eq.(\ref{eqncollapse}) obeys the well-known
rules of the quantum theory.

Separation of the quantum trajectory evolution into attraction towards the
measurement eigenstates and stochastic measurement noise exposes the striking
fact that the magnitudes of these two dynamical contributions have to be
precisely related for the Born rule to emerge as a constant of evolution.
In general stochastic processes, vanishing drift and fluctuation-dissipation
relation are quite unrelated properties, involving first and second moments
of the distribution respectively. The fact that both follow from the same
constraint ($gS_\xi=1$ or $\llangle dN\rrangle = 2g\rho_{11}dt$ in the
cases we have analysed) is an exceptional feature of quantum trajectory
dynamics. It means that the Born rule can be looked upon as a consequence
of Eqs.(\ref{fdreldiff},\ref{fdrelshot}), instead of
Eqs.(\ref{BRdiff},\ref{BRshot}). This change in view-point has powerful
implications regarding the cause of probabilistic observations in quantum
theory. Since the dissipation (convergence to the measurement eigenstates)
is produced by the system-apparatus interaction, the precisely related
fluctuations (noise giving rise to probabilistic measurement outcomes) 
too must be produced by the same system-apparatus interaction. The rest of
the environment may contribute to decoherence, but it can influence the
measurement outcomes only via the apparatus and not directly!

Another feature brought forth by our analysis is the complementary
relationship between the processes of decoherence and measurement.
An important consequence of experimental interest is to check whether
the system relaxation can be suppressed by reducing the apparatus
decoherence (or vice-versa).

Each quantum trajectory with its noise history can be associated with
an individual experimental run, and can be considered one of the many
possibilities that make up the ensemble. A model for the measurement
apparatus is needed, however, to understand where the noise comes from.
During measurement, the observed signal is amplified from the quantum
to the classical regime \cite{amplifrev}. The interactions involved are
usually electromagnetic, and often the dynamics is nonlinear. Coherent
states that continuously interpolate between quantum and classical regimes
are a convenient choice for the apparatus pointer states. They are the
minimum uncertainty (equal to the zero-point fluctuations) states in the
Fock space. {\it The crucial point is that amplification incorporates
quantum noise when the extracted information is not allowed to return}
(e.g. spontaneous vs. stimulated emission with precisely related magnitudes).
So amplifiers can indeed provide attraction towards the measurement
eigenstates together with the requisite noise. That is a direction worth
investigating further, in order to find the cause of the noise and the
irreversible collapse, and hopefully to construct a more complete theory
of quantum measurements.

The quantum trajectory framework that we have advocated does not solve
the fundamental measurement problem. What it does is to separate the Born
rule from the irreversible collapse, by explaining the system-dependent
probabilistic measurement outcomes in terms of a system-independent (but
apparatus-dependent) stochastic noise. The location of the ``Heisenberg
Cut", defining the cross-over between quantum and classical regimes, is
thus shifted higher up in the dynamics of the amplifier. This cut is not
a universal feature, but depends on the hardware of the measurement
apparatus, in terms of the type of the noise and how it originates in
the amplification process. The fluctuation-dissipation relation, and
the Born rule implied by it as per our analysis, quite likely transcend
the specific nature of the noise. It is certainly a challenge to figure
out whether the fluctuation-dissipation relation is universal for all
amplifiers, or whether it is possible to design amplifiers that would
bypass or modify the noise under some unusual conditions.

Finally, the quantum trajectory framework we have analysed can be vindicated
by verifying its predicted trajectory distributions in weak measurement
experiments. In these experiments, the coupling $g$ is a characteristic
parameter for each system-apparatus pair, and is not a universal constant.
Also, $g$ can be tuned by varying the circuit parameters without changing
the apparatus size, and it has to be made small enough to observe the
intervening stages between the initial state and the final projective outcome.
Given the type of the noise, the complete trajectory distribution (not just
its first two moments) is determined in terms of a single evolution parameter,
as evidenced by Eqs.(\ref{qdtrajdist},\ref{qjtrajdist}). The experimental
technology has developed enough for observing such trajectory distributions
in case of superconducting qubits \cite{Vijay,Murch}, and would generalise
to other quantum systems. Work in this direction is in progress.

\smallskip
\section*{Acknowledgments}

We thank Lajos Di\'osi, Nicolas Gisin and Rajamani Vijayaraghavan for
useful conversations. AP is grateful to the Institute for Quantum Computing,
Waterloo, Canada, for hospitality during part of this work. PK is supported
by a CSIR research fellowship from the Government of India.


\begin{thebibliography}{99}
\bibitem{wavefnevol}{In terms of the wavefunction, the evolution equation is
         $d|\psi\rangle/dt = g(P_i - \langle\psi|P_i|\psi\rangle)|\psi\rangle$
         with $\langle\psi|(d/dt)|\psi\rangle=0$, but the density matrix
         version is easier to analyse.}
\end{thebibliography}
\end{document}